\documentstyle[aps,twocolumn]{revtex}

\begin{document}
\title{Meissner state in finite superconducting cylinders with uniform applied
magnetic field}
\author{F. M. Araujo-Moreira}
\address{Grupo de Supercondutividade e Magnetismo, Departamento de F\'{\i }sica,\\
Universidade Federal de S\~{a}o Carlos, Caixa Postal 676, S\~{a}o Carlos SP,
13565-905 Brazil}
\author{C. Navau and A. Sanchez}
\address{Grup d'Electromagnetisme, Departament de F\'{\i }sica, Universitat
Aut\`{o}noma Barcelona 08193 Bellaterra, Barcelona, Catalonia, Spain}

\twocolumn[\hsize\textwidth\columnwidth\hsize\csname@twocolumnfalse\endcsname

\maketitle

\begin{abstract}
We study the magnetic response of superconductors in the presence of low
values of a uniform applied magnetic field. We report measurements of DC
magnetization and AC magnetic susceptibility performed on niobium cylinders
of different length-to-radius ratios, which show a dramatic enhance of the
initial magnetization for thin samples, due to the demagnetizing effects.
The experimental results are analyzed by applying a model that calculates
the magnetic response of the superconductor, taking into account the effects
of the demagnetizing fields. We use the results of magnetization and current
and field distributions of perfectly diamagnetic cylinders to discuss the
physics of the demagnetizing effects in the Meissner state of type-II
superconductors.
\end{abstract}


]

\section{Introduction}

Important intrinsic parameters of superconductors, such as the lower
critical field $H_{{\rm c1}}$ and the critical current density $J_c$, are
experimentally obtained by measuring their response to an applied magnetic
field. The procedures to obtain these parameters often rely on theoretical
approaches developed for infinite samples. When considering realistic
finite-size superconductors, important complications arise, so that these
methods fail. Even in the case of a uniform field applied to a finite
superconductor, in general it is not easy to extract information about the
intrinsic parameters of the superconductor since the magnetic response may
be strongly dependent on its shape. Demagnetizing effects appearing in
finite samples make the internal magnetic field {\bf H}={\bf B}/$\mu_0$ in
the sample different than the applied one, {\bf H}$_a$. The exact relation
between both magnetic vectors is in general unknown. As a consequence of
this indetermimation of the local magnetic field in the sample, the
estimations of $H_{{\rm c1}}$ and other parameters are very complicated.
Sometimes, this problem is treated by considering a constant demagnetizing
factor $N$, so that the magnetic field in the sample volume {\bf H} is
assumed to be related to the applied magnetic field {\bf H}$_a$ by: 
\begin{equation}
{\bf H}={\bf H}_a-N{\bf M}.
\end{equation}
This procedure is correct only for ellipsoidal-shape samples and when the
external magnetic field is parallel to one of its principal axis \cite
{chendemag}. In all other cases this equation is not valid and it becomes
very hard to find a simple relation between the magnetic vectors ${\bf H}$
and ${\bf H}_a$. Moreover, in general the field ${\bf H}$ is related with $%
{\bf H}_a$ in a different way from point to point in the superconductor.


Recently, there have been important theoretical advances in treating the
magnetic response of finite superconducting samples. The magnetic response
of finite superconductors in the critical state including demagnetizing
effects have been recently calculated for strips\cite{brandtstrip} and
cylinders \cite{brandtdisk,doyle,ieee}, following previous works on very
thin strips \cite{thinstrips} and disks \cite{thindisks}. These works deal
basically with current and field distributions in superconductors in the
mixed state, with critical-state supercurrents penetrating into the bulk of
the material. Besides, Chen {\it et al.} \cite{chendemag} have calculated
demagnetizing factors for cylinders as a function of the length to diameter
ratio and for different values of the susceptibility $\chi$ (including $%
\chi=-1$).

However, to our knowledge, there has not been done a systematic study
involving the comparison between experiments and theoretical data of the
Meissner state in superconducting cylinders. This is equivalent to ask which
are the currents that completely shield a cylindrical volume and the
resulting magnetization.

In this work we systematically investigate the magnetic response of
superconducting cylinders in the complete shielding state, by quantitatively
studying the effect of demagnetizing fields in their Meissner response. This
paper is organized as follows. In Section II we discuss the experimental
setup and measured samples. In Section III we present the experimental
results obtained from DC magnetization and AC magnetic susceptibility
techniques, performed on niobium cylindrical samples. We introduce in
Section IV a new approach to study the magnetic response of completely
diamagnetic finite cylinders. This model allows the interpretation and
understanding of our experimental results, which is discussed in Section V.
The model enables us to calculate the surface current distributions
resulting from the magnetic shielding of the cylinder, as well as the
magnetic fields created in the exterior of the samples by the induced
supercurrents. The results are discussed in Section VI. Finally, we
summarize our conclusions in Section VII.

\section{Samples and experimental setup}

We have performed the magnetic characterization of ten cylindrical niobium
polycristalline samples with different values for the ratio $L/R$, where $L$
and $R$ are the length and the radius of the sample, respectively (see Table
I). They were obtained from two different pieces of brut niobium. After
machined, they were cleaned in ultrasound bath, and later by using a HCl-HNO$%
_{3}$ solution.

We have studied two families (with five samples in each one) of cylinders
with nominal diameters of 1.94 mm and 2.86 mm, respectively. Samples with
the same nominal value for the diameter were cut from the same piece, in
cylinders of different lengths. The variation between nominal and real
values of the diameter is smaller than 4\%.

Before performing magnetic measurements, we have determined the quality of
all samples through x-ray diffraction (XRD) and scanning eletron microscopy
(SEM) analysis. For the XRD we use a SIEMENS D/5000 difractometer, and for
the SEM, we use a JEOL JSM-5800/LV microscope.


The magnetic characterization was performed through both the magnetization
as a function of the external DC magnetic field, $M(H_a)$, and the complex
AC magnetic susceptibility, as a function of the absolute temperature, $\chi
(T) $. To perform those experiments we have used a Quantum Design MPMS5
SQUID magnetometer able to operate in the ranges 2 K $<$ $T$ $<$ 400 K, 0.1
A/m $<$ $h$ $<$ 300 A/m, and 1 $<$ $f$ $<$ 1000 Hz, where $T,$ $h$ and $f$
are the absolute temperature, and the amplitude and the frequency of the AC
magnetic field, respectively. In all cases, to avoid trapped magnetic flux,
samples were zero-field-cooled (ZFC) before each experiment. The magnetic
field was always applied along the axis of the cylinders.

\section{Experimental results}

The XRD and SEM analysis confirmed the high quality of the niobium samples,
as verified by the absence of impurities and the low density of grain
boundaries. Measurements of the complex AC magnetic susceptibility, $\chi
(T) $=$\chi ^{\prime }(T)$+ $\chi ^{\prime \prime }(T)$, were performed with
the parameters $h=80$ A/m, and $f=100$ Hz, and are shown in Fig. 1. These
experiments also confirm the quality of our polycristals, by showing a
critical temperature $T_{C}=9.2$ K, and sharp superconductiong transitions,
with typical width of $0.2$ K. This value is considered excellent for a
polycristalline sample. These measurements also point out the strong shape
effect on $\chi (T)$. As can be seen in Fig. 1, the lower the ratio $L/R$
is, the larger the shape effect is, evidenced for larger values of the
modulus of $\chi ^{\prime }(T)$.

As mentioned above, we have also verified the magnetic behavior of the
cylinders by measuring isothermal $M(H_a)$ curves. The measurements were
performed at $T=8$ K. We show these results in Fig. 2. There, we can see
again the strong shape effect on the obtained curves. The lower the ratio $%
L/R$, the higher the value of the initial slope $M/H$. Also, we can see the
shape effect on the point at which the magnetization curve departs from a
straight line, which corresponds to the position where the flux starts to
penetrate in the superconductor. Thus, flux penetration starts at lower
fields for large values of $L/R$, as expected.

As it was mentioned in the last section, samples with the same nominal
diameter were cut from the same piece, in cylinders with different lengths.
Since the variation between nominal and real values of the diameter is
smaller than 4\%, only shape effects will be responsible for the observed
difference in their magnetic response.

\section{Description of the model}

In order to study the Meissner state and analyze the experimental data, we
have developed a model to calculate the surface currents that shield any
axially symmetric applied magnetic field inside a cylindrical material. This
model could be applied, in principle, to both type-I and type-II
superconductors below $H_{{\rm c}}$ and $H_{{\rm c1}}$, respectively, and
also to good conductors in a high frequency applied magnetic field.

It is well known that supercurrents are induced in the superconductors in
response to variations of the applied magnetic field. For zero-field cooled
type-II superconductors in the Meissner state, in order to minimize the
magnetic energy, these supercurrents completely shield the applied magnetic
field inside the superconductor. This shielding occurs over the whole sample
volume except for a thin surface shell of thickness $\lambda $ (the London
penetration depth), where supercurrents flow. Our model simulates this
process. It will enable us to calculate the current distribution that
shields the applied magnetic field, by finding at the currents that minimize
the magnetic energy in the system. For simplicity, we will assume that $%
\lambda $ is negligible. In this approximation, the calculated supercurrents
flow only in the cylinder surfaces.

We consider a cylindrical type-II superconductor of radius $R$ and length $L$
with its axis in the $z$ direction, in the presence of a uniform applied
magnetic field ${\bf H_{a}}=H_{a}{\bf {\hat{z}}}$. We use cylindrical
coordinates $(\rho ,\varphi ,z)$. Owing to the symmetry of the problem all
shielding currents will have azimuthal direction. We divide the
superconducting cylinder surfaces into a series of concentric circular
circuits in which currents can flow, with no limitation in the value of
current. We consider $n$ of such circuits in each one of the cylinders ends
and $m$ circuits in the lateral surface (see Fig. 3). These linear circuits
are indexed, as seen in Fig. 3, from $i=1$ to $i=2n+m-1$.

The method to obtain the current profile is the following. We start with a
zero-field cooled (ZFC) superconductor and set an applied field $H_{a}$. The
magnetic flux that threads the area closed by the $i-$circuit due to the
external field is: 
\begin{equation}
\Phi _{i}^{a}=\mu _{0}\pi \rho _{i}^{2}H_{a},  \label{extflux}
\end{equation}
$\rho _{i}$ being the radius of the $i-$circuit. The magnetic flux
contributes with an energy that has to be counteracted by induced surface
currents. A step of current $\Delta I$ that is set in some $j-$circuit
requires an energy: 
\begin{equation}
E_{j}=\frac{1}{2}L_{j}(\Delta I)^{2},
\end{equation}
being $L_{j}$ the self-inductance of the $j-$circuit, while it contributes
to reduce the energy by a factor $(\Delta I)\Phi _{j}^{a}$. So, after an
external field is applied, we seek for the circuit that decreases the energy
the most and set a current step $\Delta I$ there. The same criterion is used
next to choose either to increase the current at the same circuit by another
step $\Delta I$, or instead put $\Delta I$ at some new circuit. In the last
case (and in the general case when they are curents circulating in many
circuits of the sample), the energy cost to set a current step has an extra
term coming from the mutual inductances of all other currents: 
\begin{equation}
E_{j}=\left( {\sum_{k\neq j}}M_{kj}I_{k}\right) \Delta I_{j}.
\end{equation}
The mutual inductances $M_{ij}$ between the $i$ and $j$ linear circuits are
calculated using the Neumann formulas (see, for example, Ref. [8]). To avoid
diverging self-inductances, we have used a cut-off and calculated the
self-inductance of a circuit with radius $\rho $ from the mutual inductance
between the considered circuit and one with the same current at a radial
position $\rho +\epsilon $. An appropriate choice for $\epsilon$ has been
found to be $\epsilon = 0.78 \Delta R$, where $\Delta R$ is the radial
separation between two consecutive circuits in the end faces ($\Delta R = R/n
$).

The minimun energy correspoding to a given value of the applied magnetic
field will be reached when it becomes impossible to further decrease the
magnetic energy by setting extra current steps. From the existing current
profile, we can easily obtain the different magnitudes we are interested in.
The magnetization, which has only axial direccion $M_{z}$, is calculated by
using: 
\begin{equation}
M_{z}=\frac{1}{R^{2}L}\sum_{i}I_{i}\rho _{i}^{2}.  \label{magnetization}
\end{equation}
The magnetic induction {\bf B} could be computed from the current profiles
using the Biot-Savart formulas. However, we use a simpler way, which allows
the calculation of {\bf B} from the flux. In our model, the total magnetic
flux that threads any circular circuit (not necessarily those in the
surfaces of the superconductor) can be easily calculated as: 
\begin{equation}
\Phi (\rho ,z)=\Phi ^{a}(\rho ,z)+\Phi ^{i}(\rho ,z),
\end{equation}
where the internal flux that threads a $j-$circuit due to all the currents
is $\Phi _{j}^{i}=\sum_{k}M_{jk}I_{k}$ (the term of the self-inductances is
also included). Then, the axial component of the total magnetic induction is
simply: 
\begin{equation}
B_{z}(\rho ,z)=\frac{\Phi (\rho +\Delta R,z)-\Phi (\rho ,z)}{\pi ((\rho
+\Delta R)^{2}-\rho ^{2})}.  \label{field}
\end{equation}
The radial component $B_{r}(\rho ,z)$ is calculated from the values of the
axial one, imposing the condition that the divergence of ${\bf B}$ equals
zero.

Finally, if the external magnetic field is further increased, the same
procedure starts again from the present distribution of current.

The values of $n$ and $m$ have been chosen sufficiently large so that the
results are independent of their particular value. Typical values are $%
n\cdot m\sim 7500-10000$. The computation time for a initial curve in a
personal Digital Workstation takes few minutes for any $L/R$ ratio.

We would like to remark that, with the method described here, we obtain the
current distribution, field profiles, magnetization, and all the other
results without using any free parameter. Only the direction of currents has
to be known, which is straightfordward in this geometry.

\section{Comparison of experimental and theoretical data}

In Fig. 4 we compare the experimental values of the initial magnetic
susceptibility $\chi _{{\rm ini}}$ obtained from both DC and AC
measurements, with those calculated from our model, for different values of $%
L/R$. The calculated values of the initial slope are only function of the
length-to-radius ratio $L/R$. The agreement with the experimental data is
then very satisfactory, confirming the validity of our theoretical approach.
Both experimental and calculated data indicate a strong increase in the
absolute value of the initial susceptibility when decreasing the cylinder
aspect ratio. When the sample is very large (for $L>10R$), the initial
susceptibility $\chi _{{\rm ini}}$ approaches the value predicted for
infinite samples, $\chi _{{\rm ini}}=-1$. For shorter samples, the
magnetization gets larger in magnitude for the same value of the applied
magnetic field. We find that the general behavior of the dependence of $\chi
_{{\rm ini}}$ on $L/R$ is well described (with a departure of less than 1.5
\% from our calculations) by the approximate formula given by Brandt \cite
{brandtdisk}: 
\begin{equation}
\chi =-\pi R^{2}L-{\frac{8}{{3}}}R^{3}-{\frac{4R^{3}}{{3}}}{\rm tanh}\left[
1.27{\frac{L}{{2R}}}{\rm ln} \left( 1+{\frac{2R}{{L}}}\right) \right] .
\label{brandteq}
\end{equation}
The values of the initial slope calculated from our model are also
compatible with those given by Chen {\it et al.} \cite{chendemag} for
cylinders with $\chi =-1$ with a maximum deviation of 1.0\%.

\section{Discussion}

\subsection{Effect of the demagnetizing fields}

The experimental and theoretical results can be understood from analyzing
the effect of the demagnetizing fields. In an irrealistic infinitely long
sample, if the field is applied along its axis, shielding currents will flow
only in the lateral surface, with a constant value along the cylinder. This
creates a spatially uniform magnetic field over the superconductor, which
makes {\bf B}=0 inside. In finite samples, however, in the top and bottom
end surfaces the tangencial magnetic field is not continuous and shielding
currents are also there induced. Hence, these currents create an extra
non-uniform magnetic field over the lateral surface of the cylinder, so the
currents flowing in this surface will not have a constant value. It is
easily seen (by a simple examination of the magnetic field created by each
single current loop) that the effect that the demagnetizing fields produce
in the lateral surface region is to enhance the local magnetic field. In
this case, its total value is larger than the actual value of $H_a$. As a
result, higher values of current are necessary to shield the applied
magnetic field, yielding to larger values of both the magnetization and the
magnetic susceptibility. It is clear that, the thinner the sample is, the
larger this effect occurs, which explains the increase of $\vert\chi\vert $
when decreasing $L/R$.

\subsection{Field and current profiles}

The previous discussions can be illustrated by studying the distribution of
the magnetic field in finite cylinders. In Fig. 5 we show the calculated
total magnetic induction {\bf B} for cylinders with three different values
of $L/R$. The displayed lines indicate the direction of {\bf B} (tangential
to the lines at each point), although their densities do not reflect in
general the field strength. These results show that only in the case of the
largest $L/R$ ratio the magnetic field in the lateral surface can be
considered as basically having a constant (axial) direction over the
cylinder length. In the other two cases, the magnetic field loses its main
axial direction, gradually bending towards the axis at both cylinder ends.

In Fig. 6 we plot the calculated surface current density corresponding to
the $L/R=1$ case. In both top and bottom ends of the cylinder, the strength
of the shielding supercurrents flowing in the azimuthal direction gradually
grows (in absolute value) from a zero value over the axis to a diverging
behavior when reaching the cylinder edge (this divergence is smoothed
because of our discretization; in actual experiments the nonzero value of $%
\lambda $ makes also smooth values). The currents in the lateral surface are
also stronger at the edges and their intensity decrease towards the center
of the sample, where they have a roughly constant value over the plateau
shown in Fig. 6. The extension of this plateau (if defined as the region
where the surface current differs less than 5\% with respect to the minimum
value, located at the center) is of about 70 \% of the total cylinder
length, for $L/R=10$. This percentage decreases to 46\% and 36\% for $L/R=1$
and $L/R=0.2$, respectively (for clarity, Fig. 6 does not show the data for
the cases $L/R=0.2$ and 10). This is in correspondence with the regions,
shown in Fig. 5, where the magnetic field is almost constant. Besides, our
calculations show that the relative contribution of the currents in the end
surfaces to the magnetization increases by decreasing the cylinder
thickness. The contributions of the lateral and the end surfaces to the
total magnetization are depicted in Fig. 7. These results show that, whereas
in a long sample ($L/R=10$) about 94\% of the contribution to the
superconductor magnetization comes from the lateral surface, this percentage
decreases to about 72\% and 43\%, for $L/R=1$ and $L/R=0.2$, respectively.

\subsection{Remarks about the generality of the results}

In the $M(H_{a})$ experimental data, Fig. 2, the curves for different $L/R$
show a systematic behavior in the first portion of the initial curve (region
of interest). There, the shielding should be perfect. Nevertheless, the
trend is not so clear for larger values of the applied magnetic field. In
the middle part of the loop, it is known that bulk supercurrents penetrate
into the superconductor, which goes from the Meissner state to the mixed
state. Within the critical-state model framework, the magnetization in the
mixed state is known to be dependent on the particular $J_{c}(B)$ dependence
of the samples\cite{chengold}. This depends itself on factors such as the
detailed microstructure of the sample. Thus, samples with different
microstructures may have a difference distribution of pinning strenghts and
locations, which influences the critical current and the magnetization. The
samples we have measured were cut from two different cylinders, which
explains why the $M(H_{a})$ loops for intermediate values of $H_{a}$ are
different. The systematic behavior observed in the initial susceptibility
despite using samples from different original pieces, supports the
generality of our results for any diamagnetic cylinder.

\section{Conclusions}

In this work we propose a new model based on energy minimization which
allows the calculation of the magnetic response for perfectly diamagnetic
cylinders of any size with high precision. The only assumption we have
considered is that the magnetic penetration depth, $\lambda$, is negligible.
We have experimentally verified the validity of this model by measuring the
low-field magnetic response of niobium cylinders with different values for
the length-to-radius ratio. We have demonstrated both experimentally and
theoretically that the value of the initial susceptibility of zero-field
cooled type-II superconductor cylinders is a function of only the sample
aspect ratio, and calculated its value for a wide range of sample
dimensions. The model is sufficiently general to be adapted to other
geometries and also to non-uniform axially symmetric applied fields. These
results may help to discriminate whether some of the effects recently
observed in the study of thin film high-$T_c$ superconductors are due to
intrinsic causes or instead have an extrinsic origin associated with sample
size effects.

\section*{Acknowledgements}

We thank DGES project number PB96-1143 for financial support. C.N.
acknowledges a grant from CUR (Generalitat de Catalunya). FMAM gratefully
acknowledges financial support from brazilian agencies CNPq and FAPESP,
through grants 98/12809-7 and 99/04393-8.

\begin{table}[tbp]
\caption{Values of length (L) and radius (R) for the ten Nb samples studied}
\begin{tabular}{cccc}
Samples & L (mm$\pm$ 0.01) & R (mm$\pm$ 0.01) & Measurements \\ 
A1 & 9.72 & 1.43 & AC and DC \\ 
A2 & 1.44 & 1.43 & DC \\ 
A3 & 0.30 & 1.39 & DC \\ 
A4 & 1.38 & 1.43 & AC \\ 
A5 & 0.3 & 1.43 & AC \\ 
B1 & 9.82 & 1.00 & AC and DC \\ 
B2 & 1.50 & 0.97 & DC \\ 
B3 & 0.30 & 0.96 & DC \\ 
B4 & 1.34 & 0.89 & AC \\ 
B5 & 0.44 & 0.99 & AC
\end{tabular}
\end{table}


\begin{figure}[tbp]
\caption{Real part of the AC magnetic susceptibility as a function of
temperature, $\chi ^{\prime }(T)$, performed with the parameters $h=80$A/m,
and $f=100$Hz. Samples identifications are listed in Table 1. }
\end{figure}

\begin{figure}[tbp]
\caption{DC Magnetization loops as a function of the applied magnetic field, 
$M(H_a)$, at $T=8$ K. Samples identifications are listed in Table 1.}
\end{figure}

\begin{figure}[tbp]
\caption{Sketch of the discretization of the cylinder used in the model.}
\end{figure}

\begin{figure}[tbp]
\caption{Initial susceptibility for diamagnetic cylinders of radius $R$ and
length $L$ as a function of $L/R$. Squares and circles are experimental data
from AC and DC experiments performed on niobium cylinders, respectively. The
line corresponds to our model calculation. Vertical error bars are smaller
than symbol size.}
\end{figure}

\begin{figure}[tbp]
\caption{Theoretical sketch of the total magnetic field lines {\bf B} for
three diamagnetic cylinders with length-to-radius ratios $L/R$= (a) 0.2, (b)
1 and (c) 10. The lines describe the field direction at every point, but the
field strength is not given by the density of lines.}
\end{figure}

\begin{figure}[tbp]
\caption{Theoretical values of the surface shielding current $K$ for the
case $L/R$=1, as a function of radial and axial positions $\rho$ and $z$,
respectively. $K$ is normalized to the value $K_0$ defined as the surface
current value that would shield the same applied field $H_a$ for an infinite
sample, that is, $K_0=H_a$. The left part of the figure corresponds to the
current in the end surfaces of the cylinder $K_{{\rm faces}}$ and the right
one to the current in the lateral surface $K_{{\rm lateral}}$. The arrows
mark the plateau of almost costant current (see text).}
\end{figure}

\begin{figure}[tbp]
\caption{Theoretical contribution from the lateral and end surfaces of
diamagnetic cylinders to the magnetization, for the cases $L/R$=0.2, 1, and
10. Dotted, dashed and solid lines correspond to the contribution from the
lateral surface, the contribution from the end surfaces, and the total,
respectively.}
\end{figure}

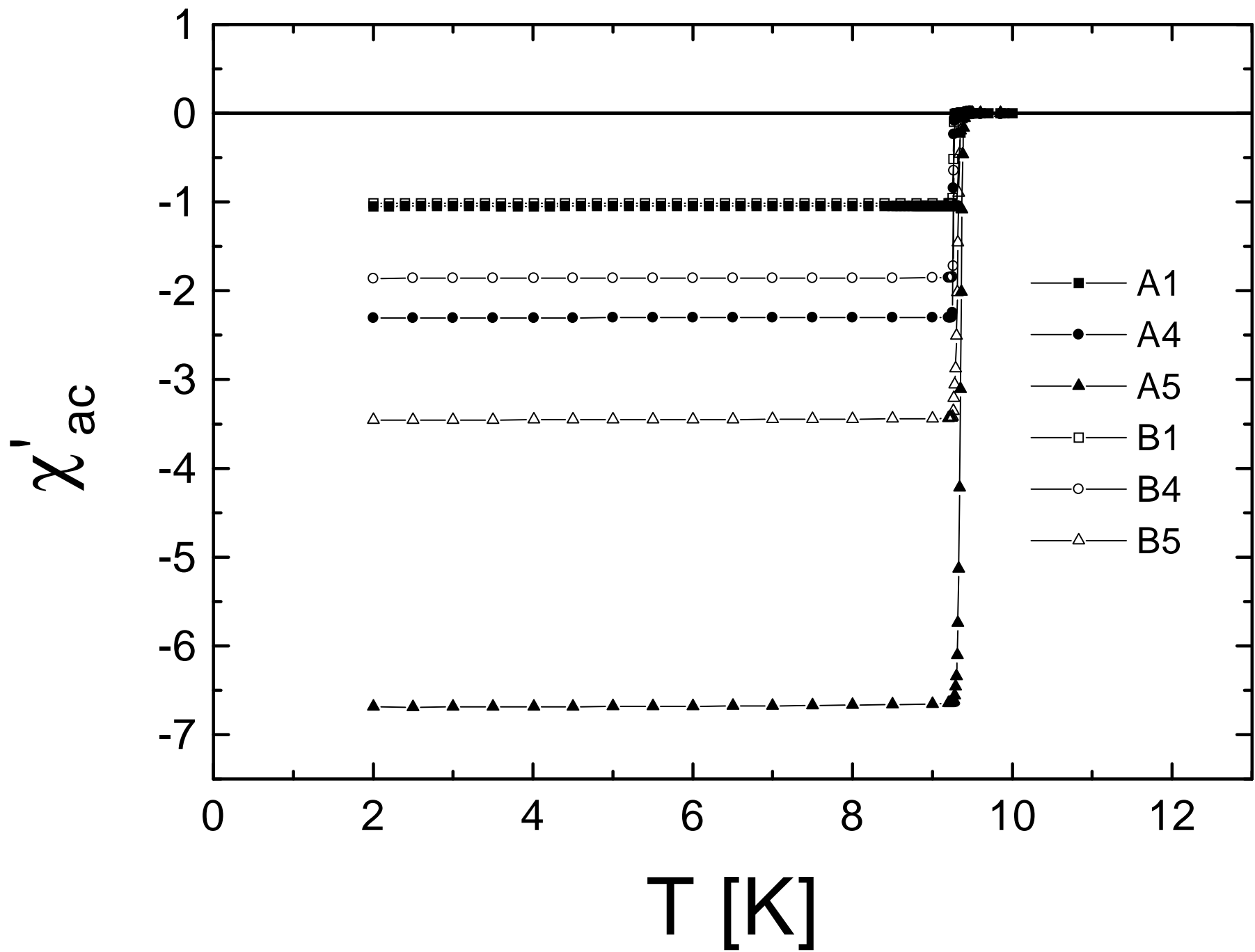
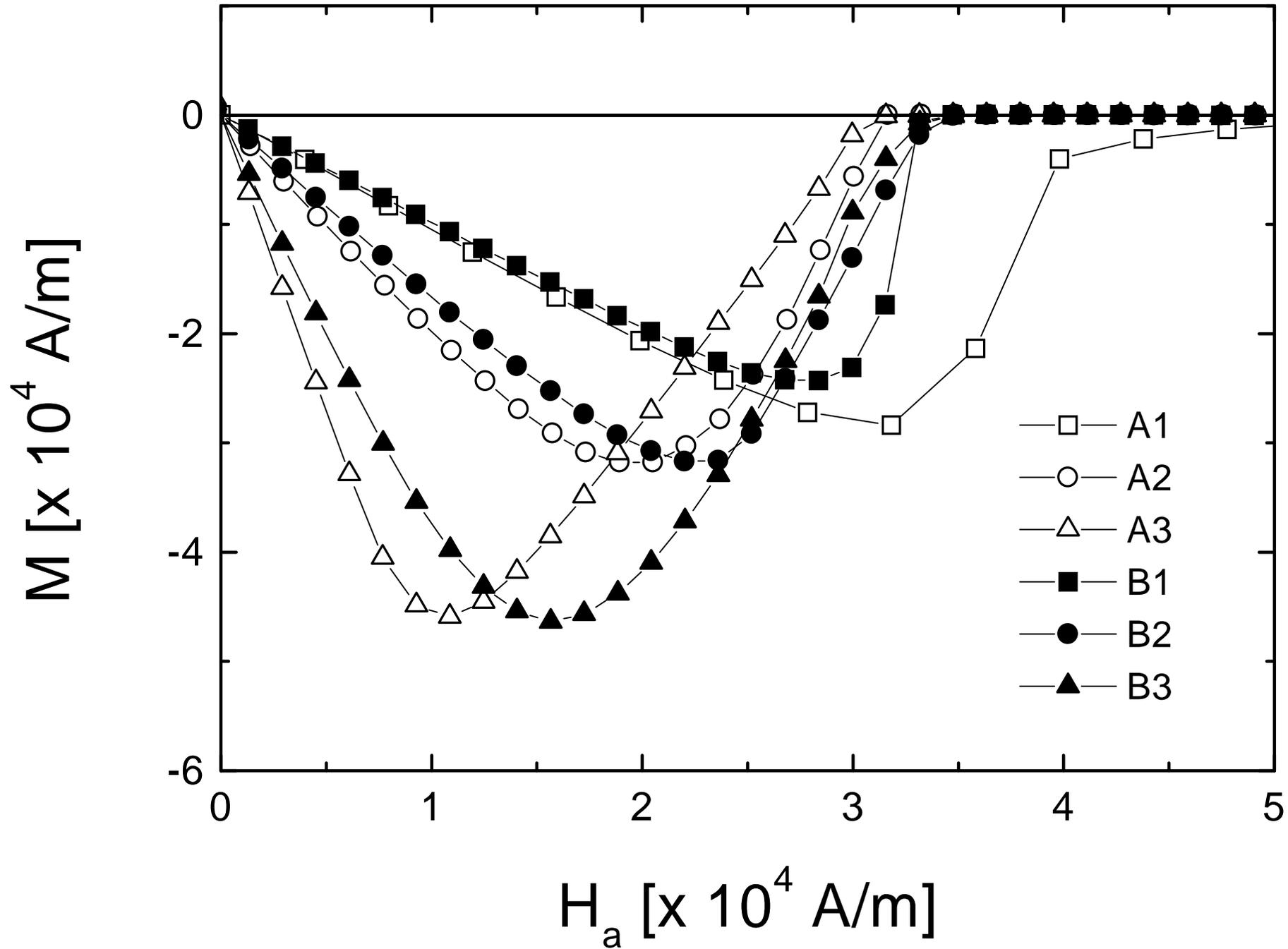
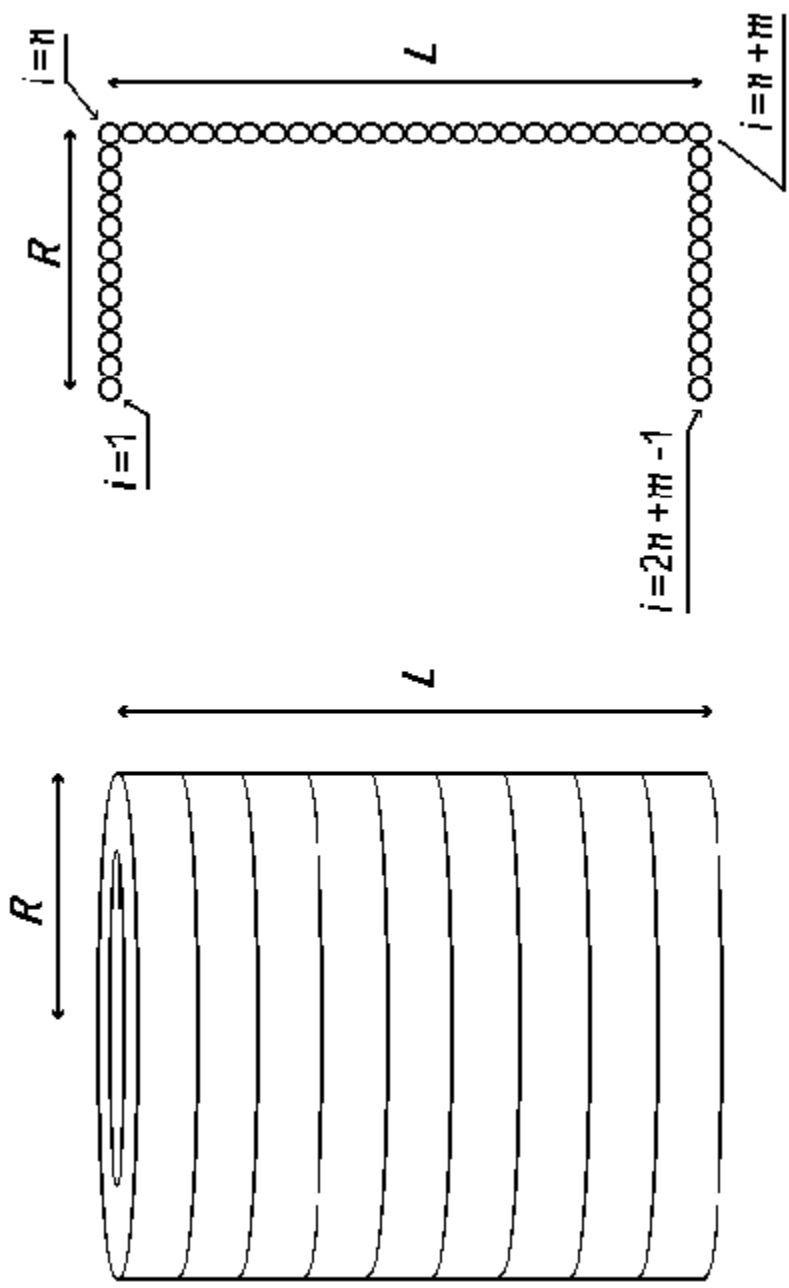
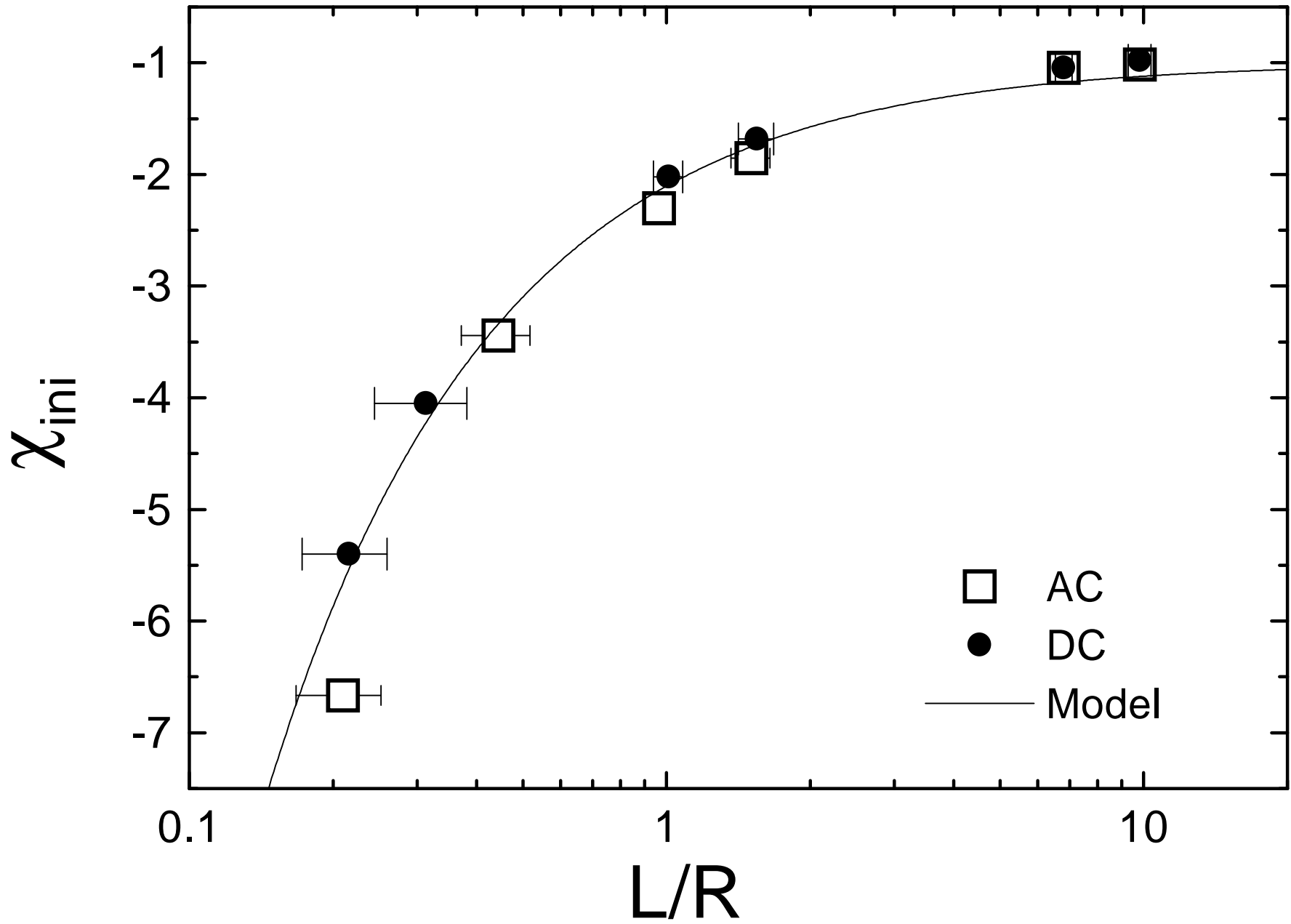
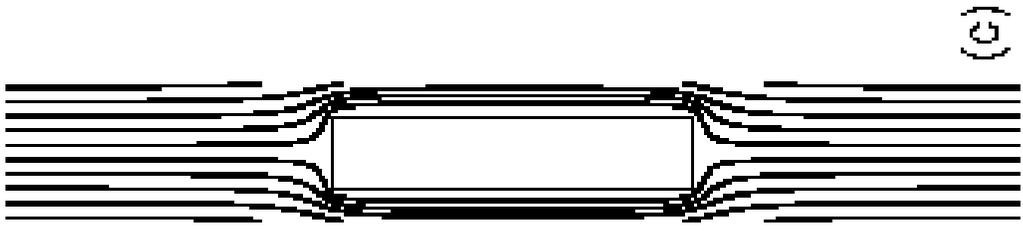
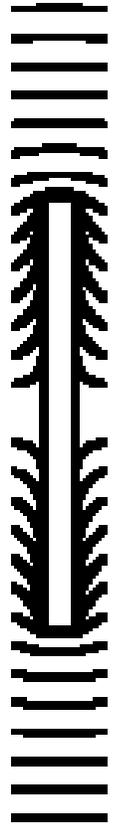
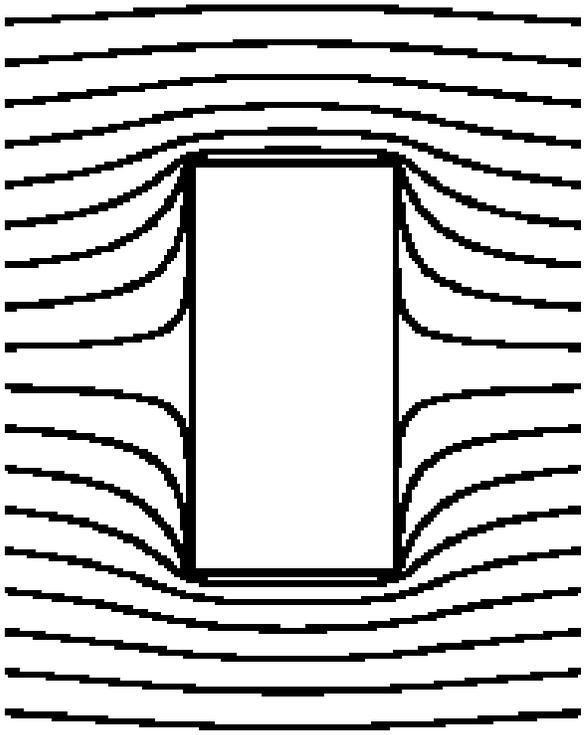
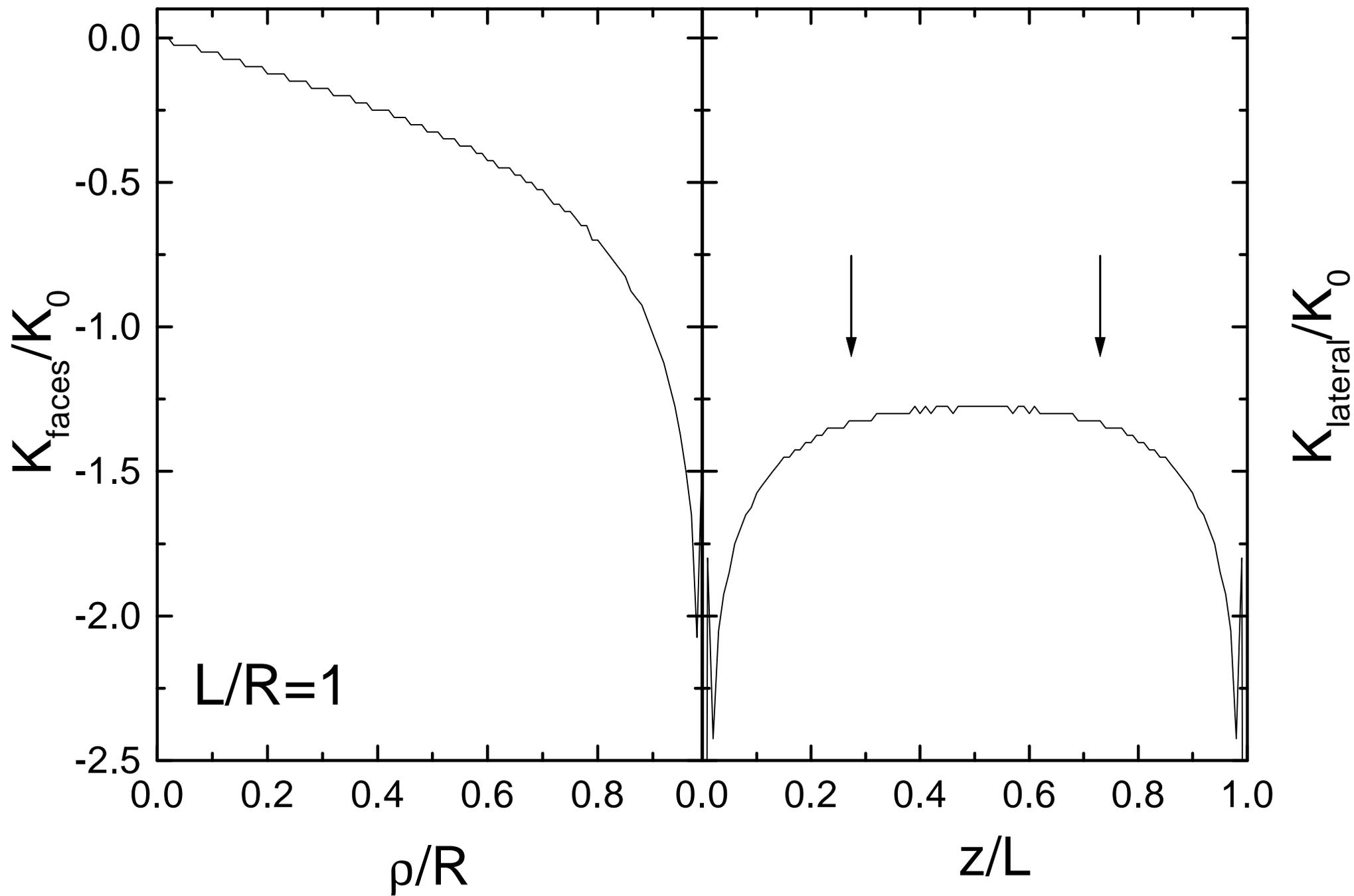
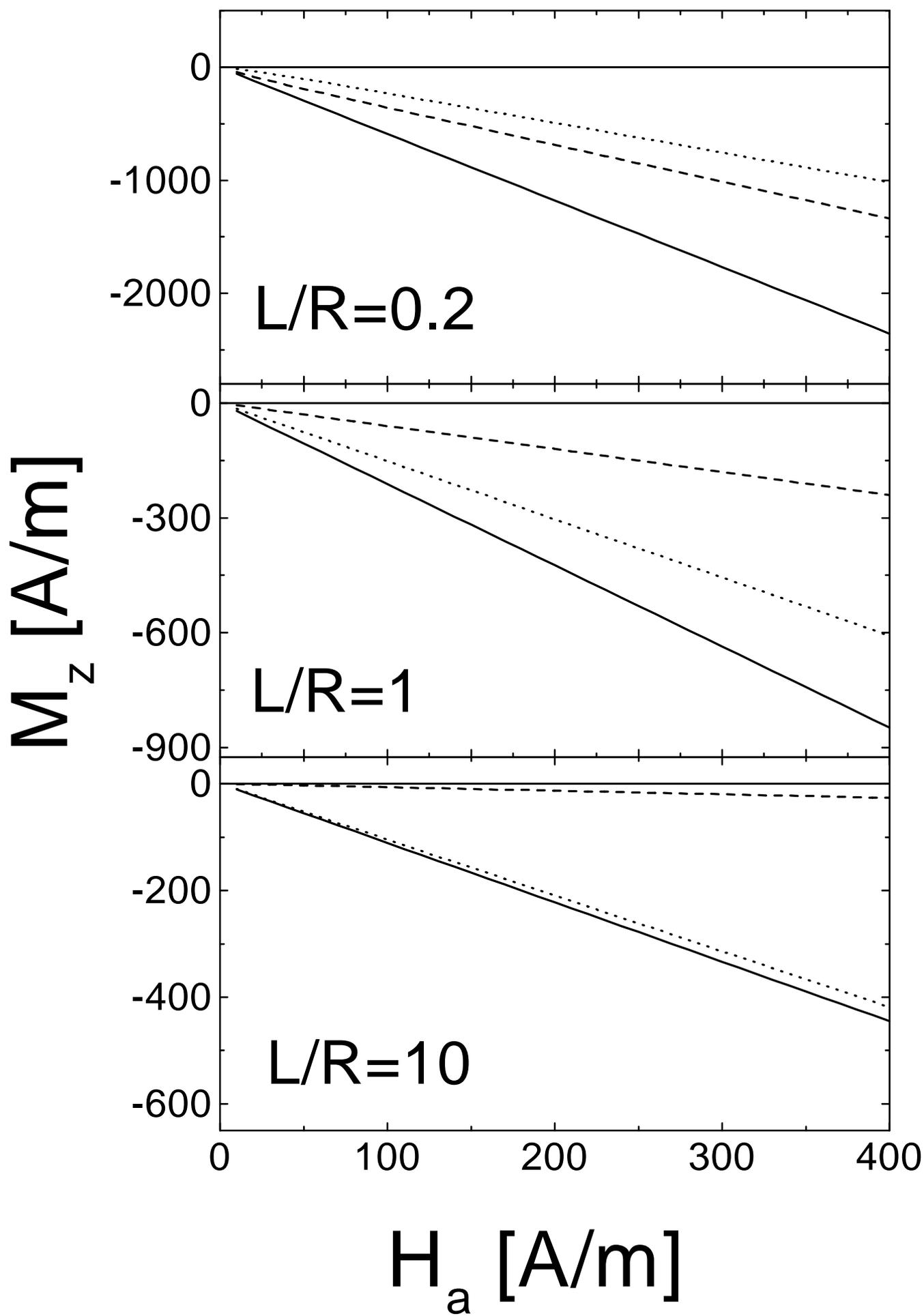
\end{document}